\documentclass[fleqn,usenatbib]{mnras}

\usepackage{newtxtext,newtxmath}

\usepackage[T1]{fontenc}
\usepackage{ae,aecompl}

\usepackage{graphicx}	
\usepackage{amsmath}	
\usepackage{amssymb}	
\usepackage{}

\title[Photoevaporation on disc `isochrones']{Effects of photoevaporation on protoplanetary disc `isochrones'}

\author[A. Somigliana et al.]{
Alice Somigliana,$^{1}$\thanks{E-mail: alice.somigliana@studenti.unimi.it}
Claudia Toci,$^{2}$\thanks{E-mail: claudia.toci@inaf.it}
Giuseppe Lodato$^{1,2}$
Giovanni Rosotti$^{3}$ and
\newauthor
 Carlo F. Manara$^{4}$
\\
$^{1}$Dipartimento di Fisica, Universit\'a degli Studi di Milano, 
Via Giovanni Celoria 16, I-20133 Milano, Italy \\
$^{2}$INAF - Osservatorio Astronomico di Brera, Via Brera 28, 20121 Milan, Italy\\
$^{3}$Leiden Observatory, Leiden University, P.O. Box 9531, NL-2300 RA Leiden, the Netherlands\\
$^{4}$European Southern Observatory, Karl-Schwarzschild-Strasse 2, 85748 Garching bei München, Germany
}
\date{Accepted 20/11/19. Received 14/10/19}

\pubyear{2019}

\begin{document}
\label{firstpage}
\pagerange{\pageref{firstpage}--\pageref{lastpage}}
\maketitle

\begin{abstract}
Protoplanetary discs are the site of star and planet formation, and their evolution and consequent dispersal deeply affect the formation of planetary systems.
In the standard scenario they evolve on timescales $\sim$~Myr due to the viscous transport of angular momentum.   
The analytical self-similar solution for their evolution predicts also specific disc isochrones in the accretion rate - disc mass plane.
However, photoevaporation by radiation emitted by the central star is likely to dominate the gas disc dispersal of the innermost region, introducing another (shorter) timescale for this process.
In this paper, we include the effect of internal (X and EUV) photoevaporation on the disc evolution, finding numerical solutions for a population of protoplanetary discs. 
Our models naturally reproduce the expected quick dispersal of the inner region of discs when their accretion rates match the rate of photoevaporative mass loss, in line with previous studies.
We find that photoevaporation preferentially removes the lightest discs in the sample. The net result is that, counter-intuitively, photoevaporation increases the average disc mass in the sample, by dispersing the lightest discs. At the same time, photoevaporation also reduces the mass accretion rate by cutting the supply of material from the outer to the inner disc. In a purely viscous framework, this would be interpreted as the result of a longer viscous evolution, leading to an overestimate of the disc age. Our results thus show that photoevaporation is a necessary ingredient to include when interpreting observations of large disc samples with measured mass accretion rates and disc masses.
Photoevaporation leaves a characteristic imprint on the shape of the isochrone. Accurate data in the accretion rate - disc mass plane in the low disc mass region therefore give clues on the typical photoevaporation rate. 
\end{abstract}




\section{Introduction}
Planets form in protoplanetary discs.
Therefore, understanding disc evolution and deepening our knowledge on the processes that are responsible for disc dispersal is essential in order to understand the formation of planetary systems. 
In particular, the processes responsible for disc dispersal are not completely understood at the present time. However, the associated time-scale sets a limit on the time-scale for gas-giant planets formation and influences the initial architecture of exoplanetary systems \citep{Alexander2012,Ercolano2015,Jennings2018}. 

The formation of protoplanetary discs is a consequence of angular momentum conservation during star formation.
Such discs evolve by removing/transporting angular momentum through the disc due to viscous stresses, although the mechanism leading to this viscosity is still debated \citep{Armitage}, with observational studies having excluded that the viscosity is very high \citep{Flaherty2017}. Viscous accretion disc theory \citep{ShakuraSunyaev1973,LyndenBellPringle1974} 
predicts a lifetime for discs of $\sim$ few Myr,
broadly consistent with observational results \citep{Fedele2010,Ansdell2017}.
However, the transition from disc-bearing to disc-less status seems to be rapid compared with the viscous theory expectation \citep{Ercolano2011a}.
In particular, 'transitional discs' can provide further information regarding the mechanism for clearing discs \citep{Strom1989,Alexander2014}. These objects show a deficit in the opacity observed in near IR (NIR) wavelenght compared with the 'standard' emission of discs, expected to be optically thick, but are consistent with theoretical optically thick disc emission at Mid-Far IR (MIR, FIR). This is usually interpreted as
a decrease of warm dust in their innermost regions, $\sim$ few au \citep{Skrutskie1990,Espaillat2010,Espaillat2014}. 
The frequency of 'transition' versus 'standard' discs suggests that $\sim 10-20 \%$ of all the discs are in this stage \citep{Luhman2010}. If all discs encounter the transition disc phase, this
means that this phase must be short-lived ($\sim 10^5$ yr),
implying that disc evolution follows a 'two-time scale behaviour', as first suggested by \cite{ClarkeGendrinSoto2001}.
Two main mechanisms have been proposed to explain the observational (in dust and gas) presence of transition discs: internal photoevaporation (see e.g \citealt{Hollenbach1994,ClarkeGendrinSoto2001,Alexander2006}) and planet formation (see e.g \citealt{Armitage1999,Rosotti2013}). 
\cite{ClarkeGendrinSoto2001} showed that considering photoevaporation together with viscous evolution leads to the opening of a cavity at around few au when the mass accretion rate and the mass loss by photoevaporation are comparable. As a consequence, the accretion rate dramatically decreases (the
so-called UV switch) and the inner disc is accreted on its own (shorter) timescale, producing a transition disc in $\sim 10^5$ yr, in agreement with observational results \citep{Ercolano2011b}. 
In recent years, attention has been paid also to other heating radiation fields such as the far ultra-violet
(FUV) \citep{GortiHollenbach2008} and soft X-rays \citep{Alexander2005,Ercolano2008,Owen2010,Picogna2019}, leading to the development of multiple photoevaporation models (see \citealt{Alexander2014} for a review) that differ in the mass-loss rates, the shape of the mass-loss profile and the link between the properties of the wind and the ones of the central star (such as the stellar UV or X-ray flux). In general, however, the more recent models tend to predict significantly higher
mass-loss rates compared to the original EUV model. There is now direct, observational evidence of the presence of slow-moving, mildly supersonic winds thanks to the detection of
blue-shifted forbidden emission lines such as [NeII], for example in TW Hydra \citep{Pascucci2011}. Photoevaporation is a natural candidate for this type of winds and the different photoevaporation models are consistent with these observations \citep{Alexander2008,ErcolanoOwen2010}. However, this also means that these observations are not capable to discriminate between the existing models, which at the moment still remain viable.

Given the existence of different, competing models,  multiwavelength observational surveys of large and complete samples can be an useful testbed to test such evolutionary models with a statistical approach,  comparing them with synthetic populations. Mass and accretion rates of discs in the star forming regions Lupus and Chamaeleon I have been measured in spectroscopic surveys \citep{Alcala2014,Alcala2017,Manara2016b,Manara2017} and in mm-interferometric surveys \citep{Ansdell2016,Pascucci2016}, and a correlation between these two quantities was observed \citep{Manara2016a,Mulders2017}.
The first analysis has been done by \cite{Hartmann1998}, that estimated the value of the viscosity studying the decreasing of the measured mass accretion rate as a function of time. Subsequently \cite{Jones2012} found that, in order to reproduce observational disc quantities such as the mass, the mass accretion rate and the ages of a sample of discs, significant deviation from the  \cite{LyndenBellPringle1974} self similar solution are needed. However, the sample was inhomogeneus and with large uncertainties in the measured quantities. Recently \cite{Rosotti2017} expanded this study considering also other mechanisms such as dead zones and photoevaporation. \cite{Lodato2017} introduced the concept of disc 'isochrones' for protoplanetary discs, that are the locus of the mass accretion rate ($\dot{\rm{M}}$) - disc mass ($\rm{M}_d$) plane where sources of same age are located. The comparison between the isochrones and a (synthetic or observed) population can give an estimate of the age or the viscosity or the initial mass distribution of a given population. In that work the authors found an analytical expression for the isochrones in the case of self-similar solutions. Considering the evolution of a population of discs with similar ages, they also showed that the correlation between $\dot{\rm{M}}$  and $\rm{M}_d$ and its scatter are functions of time, and in particular that the slope is shallower then linear, in agreement with observational results of Lupus survey \citep{Manara2016a} for young samples, while the slope approaches unity for old disc populations. \citet{manara2019} compared the disc population used in planet population synthesis models \citep{Mordasini2009,Mordasini2012} (that include EUV photoevaporation) with the data in Lupus and Chamaleon star formation region.

In this paper, we perform an analysis similar to \cite{Lodato2017}, by numerically evaluating the disc 'isochrones' for a disc population evolving through viscous evolution and internal photoevaporation.
We explore two different scenarios: a EUV-driven wind following \cite{ClarkeGendrinSoto2001} and a X-ray wind as \cite{Owen2011}.
In Sec. \ref{sec:methods} we describe the theoretical model and the numerical setup. Our results are described in Sec. \ref{sec:results}. In Sec. \ref{sec:conclusion} we discuss the presented results and we give the conclusions of this work.

\section{Method}\label{sec:methods}

The evolution of the surface density of a protoplanetary disc $\Sigma (R, t)$ can be studied as a function of the radius $R$ and the time $t$. It evolves with time following a diffusion equation, 

\begin{equation}
    \frac{\partial \Sigma}{\partial t} = \frac{3}{R} \frac{\partial}{\partial R} \left( R^{1/2} \frac{\partial}{\partial R} \left( \nu \Sigma R^{1/2}\right) \right) - \dot{\Sigma}_{\rm{wind}}.
    \label{eq:dsigma/dt_photo}
\end{equation}

Eq. \ref{eq:dsigma/dt_photo} is the most general form of the evolution equation in the case of protoplanetary discs subject to both viscosity and photoevaporation. In particular, $\dot{\Sigma}_{\rm{wind}}$ is the photoevaporative term, that accounts for the mass loss due to photoevaporation. The analytic form of this latter term depends on the theoretical model considered. In this work, we use two different parametrisations for $\dot{\Sigma}$, both due to internal photoevaporation: the analytic function of the EUV photoevaporation from \cite{ClarkeGendrinSoto2001} and the polynomial approximation of the X photoevaporation term from \cite{Owen2011}.

The first term on the RHS of Eq. \ref{eq:dsigma/dt_photo}, without the photoevaporative term, is the general equation for viscous evolution.
Assuming for the kinematic viscosity $\nu$ a power-law dependence on radius, $\nu = \nu_c (R/R_c)^{\gamma}$ where $R_c$ is a scale radius, $\gamma$ is a free index, $\nu_c$ is the value of $\nu$ at radius $R_c$ and is evaluated using the $\alpha$ prescription \citep{ShakuraSunyaev1973}, this equation can be solved analytically. The solution, called 'self-similar' solution \citep{LyndenBellPringle1974}, naturally introduces a timescale for the lifetime of the disc, the viscous time:

\begin{equation}
    t_{\nu} = \frac{R_c^2}{3(2 - \gamma) \nu_c};
    \label{eq:tnu}
\end{equation}

typically, for $t >> t_{\nu}$ discs can be considered to have reached the self-similar condition. 

Our goal in this paper is to study the isochrones for a sample of protoplanetary discs subject to photoevaporation, expanding on the work of \citet{Lodato2017} who focused on the case of purely viscous evolution. We remark here the dependence of $t_\nu$ on the value of $\alpha$, in particular according to the $\alpha$ prescription , $\nu(R)=\alpha c_s^2\Omega(R)^{-1}$, where $c_s$ is the local sound speed and $\Omega(R)$ the Keplerian angular frequency.

\subsection{Photoevaporation}

Photoevaporation occurs when radiation, coming either from the central star (internal) or from nearby massive stars (external), ionises the gas in the upper layers of protoplanetary disc: part of the mass in the disc then becomes unbound from the central star and leaves the system flowing in a thermal wind. As a result, we observe the formation of a cavity inside the disc. A big role is played by the mass-loss rate $\dot{\rm{M}}_{\rm{wind}}$, that depends on the features of the ionising radiation. The effects of photoevaporation become dominant once the accretion rate through the disc becomes low enough to be comparable with $\dot{\rm{M}}_{\rm{wind}}$ \citep{ClarkeGendrinSoto2001}. Most of the photoevaporative models find that the mass loss comes from the discs surface at radii beyond the 'gravitational radius'

\begin{equation}
    R_g = \frac{G M_*}{c_s^2},
    \label{eq:rg}
\end{equation}

where $M_*$ is the stellar mass and $c_s$ is the speed of sound in the photoionised gas; $R_g$ is therefore a measure of the radius at which the cavity initially forms. Once the cavity is open, the inner disc is  accreted onto the central star at its new own viscous timescale, which is much smaller than the original one (typically $10^5$ yr versus $10^6 - 10^7$ yr), while the outer disc is dispersed on timescales comparable to the original one: viscous evolution with photoevaporation is therefore a two-timescales process. 

In this work we followed two different photoevaporative models, respectively by \citet{ClarkeGendrinSoto2001} and \citet{Owen2012}, taking into account UV and X photoevaporation.  
Following the work of \citet{ClarkeGendrinSoto2001}, we adopted the analytical prescription for the UV photoevaporation term $\dot{\Sigma}$, with the same value of $\dot{\rm{M}}_{\rm{wind}}=4\times10^{-10} \rm{M}_\odot \rm{yr}^{-1}$. 
In the case of X photoevaporation, \cite{Owen2012} introduce an approximated polynomial prescription, numerically determined in \cite{Owen2011}. In their work, the value of $\dot{\rm{M}}_{\rm{wind}}$ ranges from $10^{-11} \rm{M}_\odot \rm{yr}^{-1}$ to $10^{-7} \rm{M}_\odot \rm{yr}^{-1}$ according to the mass and X-ray luminosity of the protostar. We chose as our reference value of the mass-loss rate the intermediate value of $10^{-8} \rm{M}_\odot \rm{yr}^{-1}$.
For a detailed discussion on the theoretical models, we refer to the original papers.

\subsection{Numerical model}

\begin{figure}
	\includegraphics[width=\columnwidth]{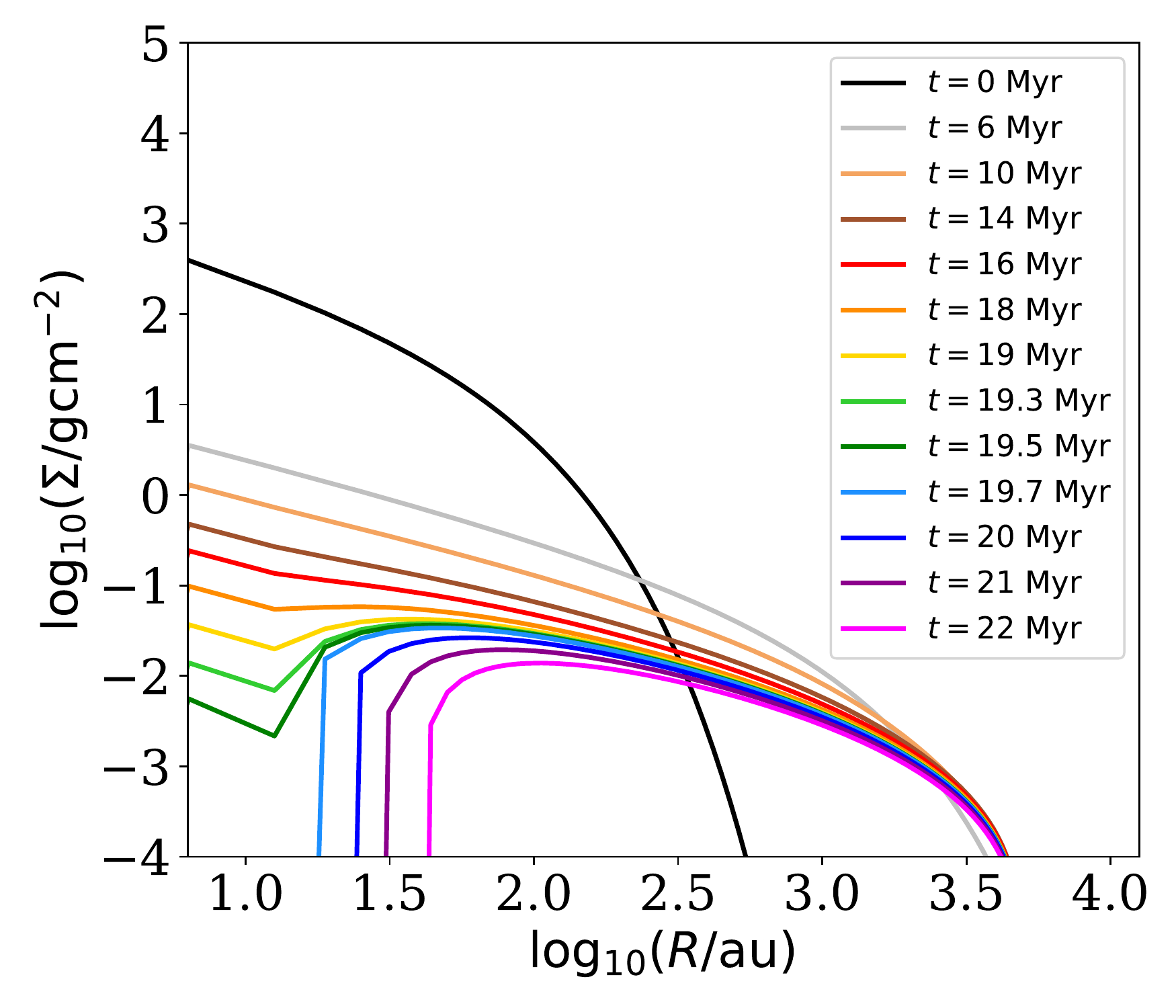}
    \caption{Plot of the surface density $\Sigma$ as a function of radius $R$ at various ages. We notice the formation of a cavity inside the disc at $R \sim 10$ au (i.e. $R = R_g$), in correspondence with the value of $R_g$. For this plot, we set $\dot{\rm{M}}_{\rm{wind}} = 4 \times 10^{-10} \rm{M}_{\odot}/ \rm{yr}$.}
    \label{fig:clarke_apertura}
\end{figure}

The starting point of our work is to create a synthetic population of protoplanetary discs to use as initial conditions for the one-dimensional diffusion code that we developed. Our code integrates Eq. \ref{eq:dsigma/dt_photo} using standard finite-difference methods \citep{num1}, and simulates therefore the discs evolution with time accounting for both viscous and photoevaporative effects. We used a radial grid of 800 meshes, equally spaced in $R^{1/2}$ with $R_{\text{in}} = 0.1$ au and $R_{\text{out}}  = 5000$ au. 
Particular attention should be reserved on the boundary values of the grid. Indeed, a small value of $R_{\text{out}}$ can introduce some mass loss out of the boundary radii for large values of $R_c$ and long time evolution, which might modify the isochrones, see Sec. \ref{sec:results}.

We generated an initial disc population spanning different values of their initial mass $M_0$ and characteristic radius $R_c$; the resulting viscous time $t_{\nu}$ is linearly proportional with the value of $R_c$ for fixed values of $\alpha$.

Rather than randomly drawing the values of the free parameters, we chose to use a set of pre-defined values to better isolate their influence on our model.
This led us to have some recurring shapes in our plots, corresponding to discs with same initial mass or typical radius (we will underline these shapes in the subsequent section). 
In order to span realistic values, we based our parameter ranges on recent observational results \citep{Manara2016a,Ansdell2017}, linearly spanning  $\log_{10}(M_0/\rm{M}_{\odot}) \in [-3, -1]$ and $\log_{10}(R_c / \rm{au}) \in [0, 3]$.  

In the following, we will always consider the case where $\gamma = 1$ and a value of the aspect ratio of the disc (at $R=R_c$) $H/R = 0.08$. 
The value of the mass-loss rate $\dot{\rm{M}}_{\rm{wind}}$ is constant and to be set, as well as the photoevaporative radius $R_g$ (which affects only the EUV case) and the value of the parameter $\alpha$. Unless otherwise stated we have always considered the EUV photoevaporation case, setting therefore $R_g = 5$ au following recent studies (see i.e \citealt{Ercolano2017}), and the value of $\alpha$ is fixed at $10^{-3}$. 
The value of $\dot{\rm{M}}_{\rm{wind}}$ will vary throughout this paper, as specified.

\section{Results}\label{sec:results}

First of all, to test our numerical code, we reproduce the results found by \citet{ClarkeGendrinSoto2001}. 
Fig. \ref{fig:clarke_apertura} shows the surface density $\Sigma$ as a function of radius $R$ at different times, fixing the value of $\dot{\rm{M}}_{\rm{wind}} = 4 \times 10^{-10} \text{M}_{\odot} \rm{yr}^{-1}$. We also set $R_g = 10$ au, $M_0$ = $10^{-1} \rm{M}_\odot$ and $R_c=10$ au. We can indeed observe the formation of a cavity inside the disc at $R = R_g$ and a trend in total agreement with our reference plot (cfr. Fig. 1.a from \citet{ClarkeGendrinSoto2001}), which make us confident in our code.

\begin{figure}
    \includegraphics[width=7.8cm]{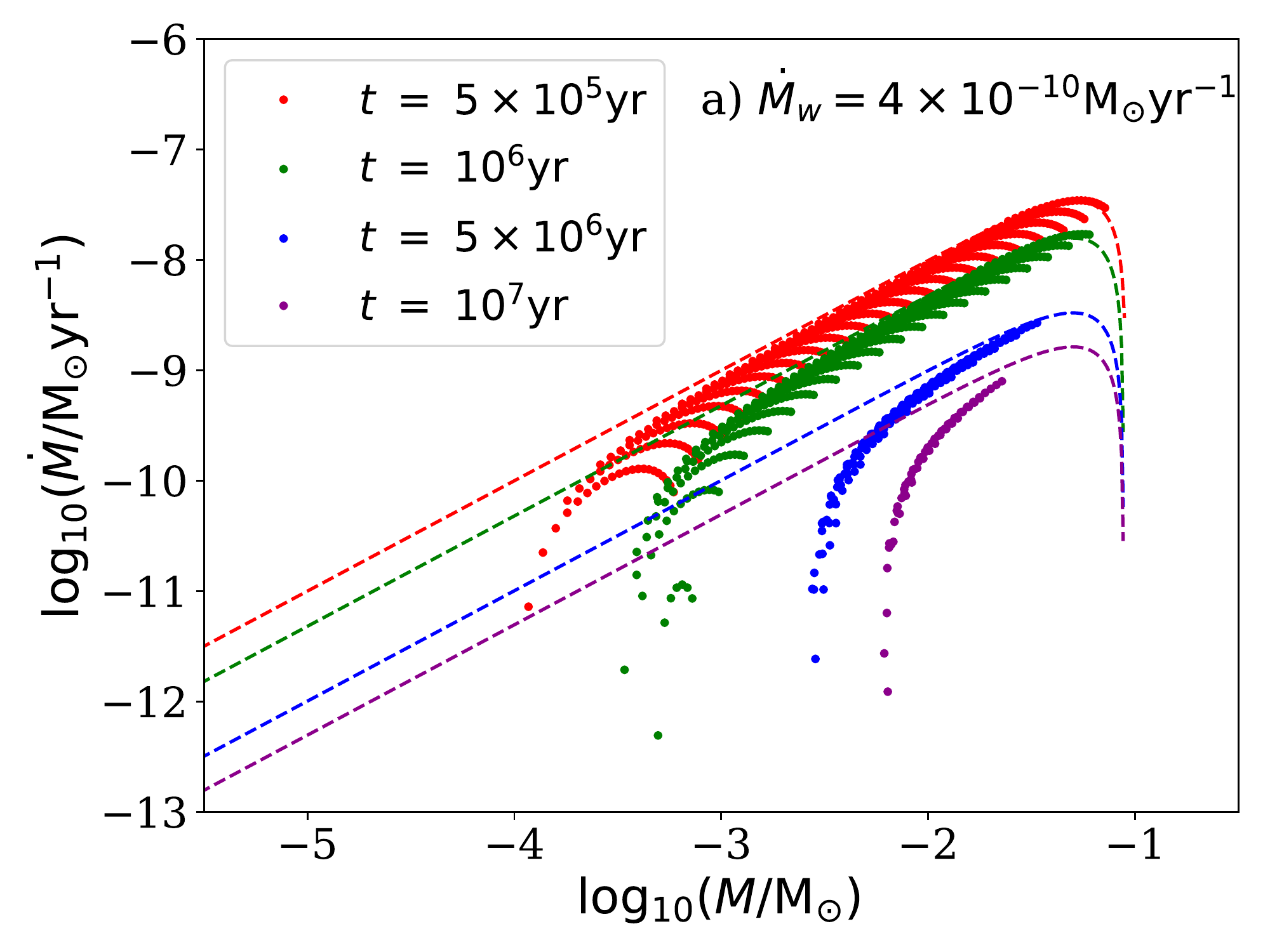}
    \vspace{0.5cm}
    \includegraphics[width=7.8cm]{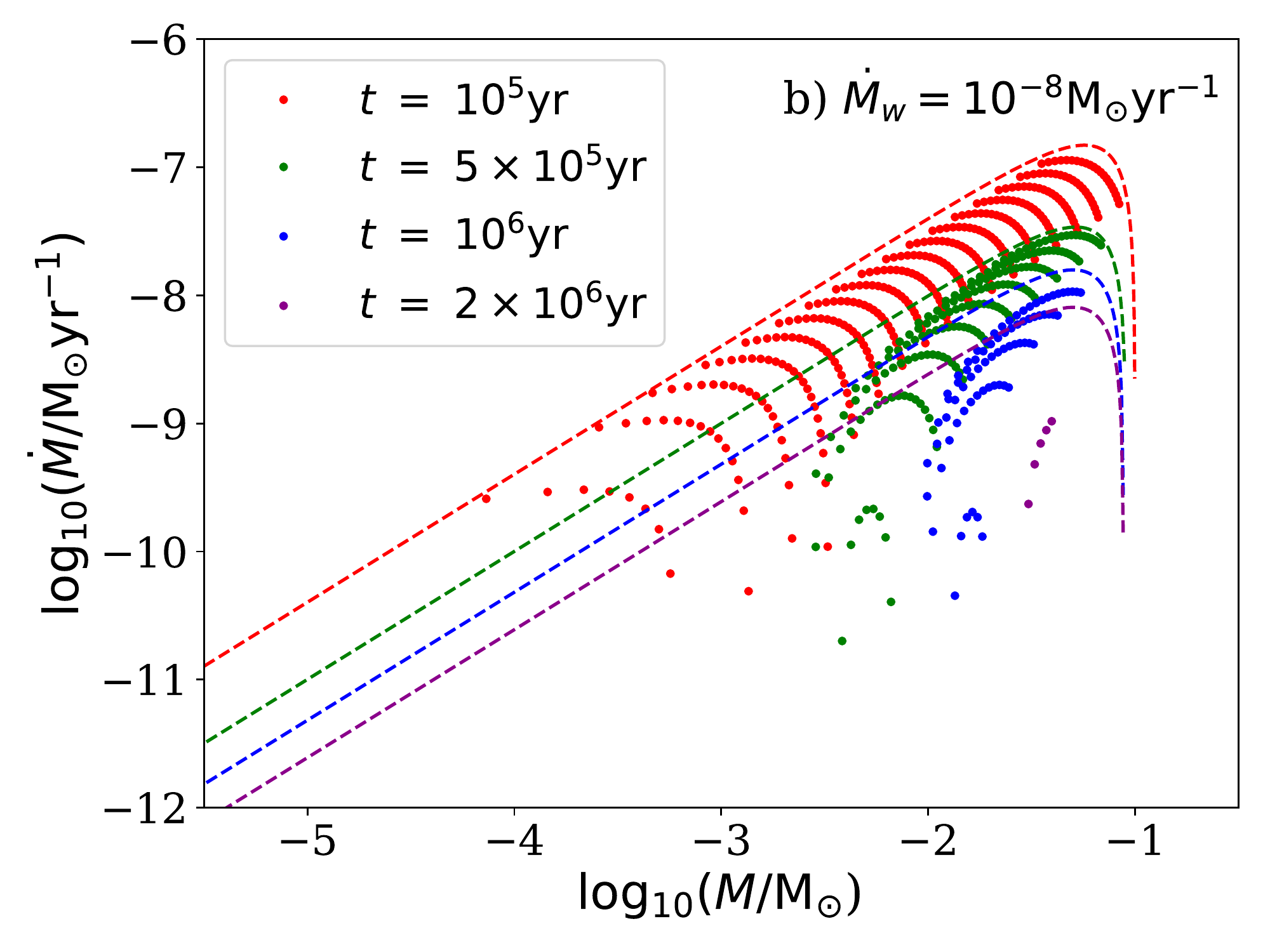}
    \caption{Isochrones for a disc population subject to photoevaporation, respectively implemented following \citet{ClarkeGendrinSoto2001} (top) and \citet{Owen2012} (bottom).}
    \label{fig:isocrone}
\end{figure}

We then focused on producing isochrones for our population for both  photoevaporative models, as shown in Fig. \ref{fig:isocrone} in the case of \cite{ClarkeGendrinSoto2001} (top) and in the case of \cite{Owen2012} (bottom). The dots represent our models, while the dashed line is the self-similar prediction at the same age. The significant difference in scatter and number of discs between the two plots is due to the mass-loss rates, chosen to be consistent with the original papers and respectively of $4 \times 10^{-10} \text{M}_{\odot} \rm{yr}^{-1}$ and $10^{-8} \text{M}_{\odot} \rm{yr}^{-1}$.

At early times, the discs position themselves on the theoretical self-similar isochrone for the corresponding initial disc mass, in line with the analytical predictions of \citet{Lodato2017}. At later times, however (and for lower disc masses and accretion rates) photoevaporation kicks in, reducing the accretion rate and therefore `bending' the isochrone downwards, as predicted by \citet{Rosotti2017}. As the population evolves, this downward knee in the isochrone moves to progressively higher disc masses. Comparing the top and bottom panels of Fig. \ref{fig:isocrone}, one can see that, as the mass-loss rate increases, the departure from the self-similar isochrone occurs at earlier times, and that the downward knee moves to higher mass for the same evolutionary time. 

The above evolution produces the counter-intuitive result that a population of photoevaporating discs is apparently \emph{more} massive than the corresponding self-similar population of the same age. This is due to the fact that the lowest disc masses have been removed because they are readily dispersed. Similarly, increasing the photoevaporative loss rate produces a population with higher disc masses on average (assuming a uniform distribution in initial disc masses and viscous timescales). We have encountered this behaviour in all the models we have run. This behaviour is the result of the competition between the rate at which photoevaporation removes low-mass discs and the rate at which high-mass discs lose their mass through accretion. Therefore, note that in principle one might get the opposite behaviour (the average disc mass \textit{decreasing} in time) by constructing initial conditions in which the most massive discs are quickly removed from the system.

\begin{figure}
	\includegraphics[width=\columnwidth]{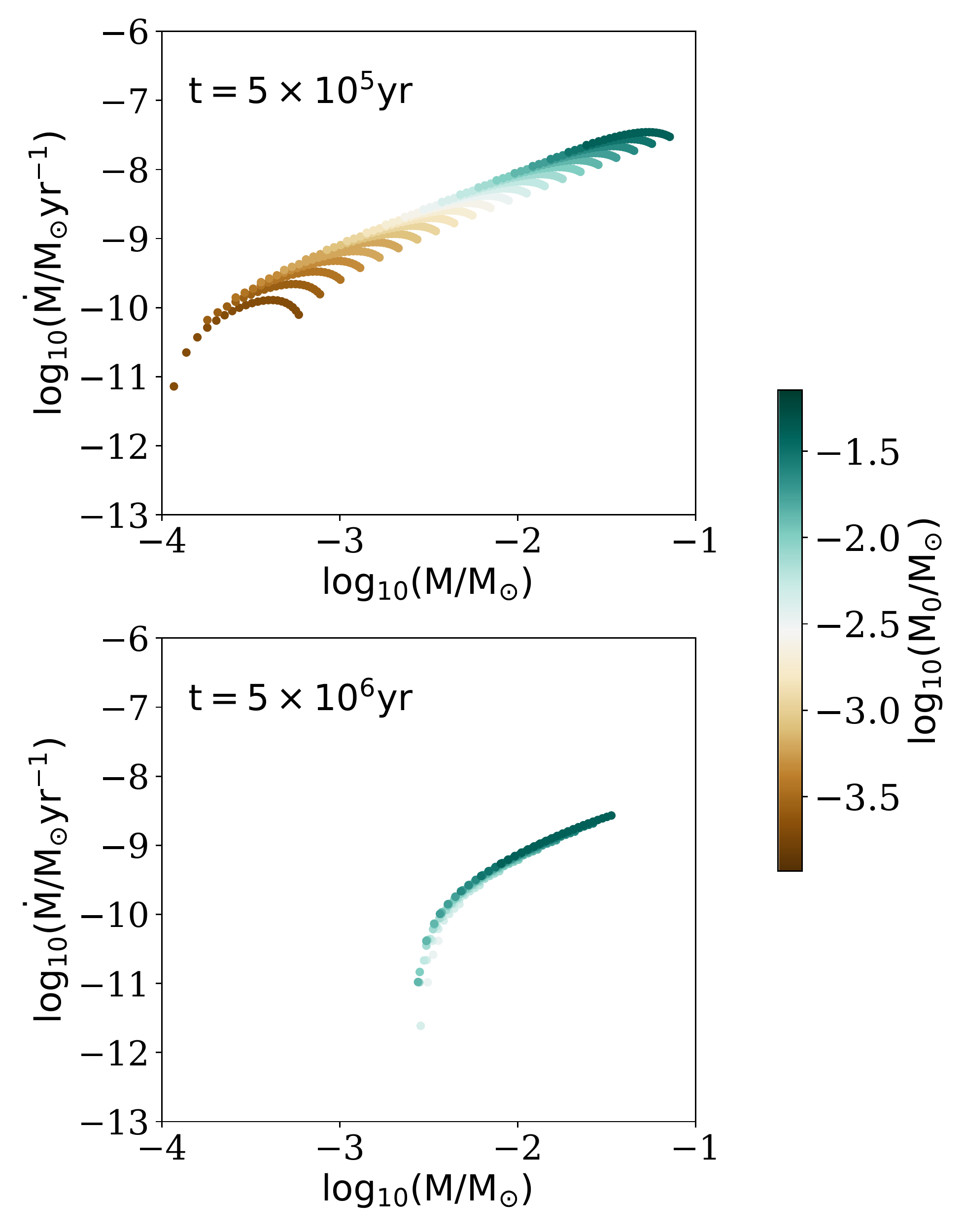}
    \caption{Isochroes for a disc population at two different ages. The colour coding refers to discs with the same initial mass $M_0$: blue dots represent discs with $M_0 = 10^{-1} \text{M}_{\odot}$, brown dots represent discs with $M_0 = 10^{-3} \text{M}_{\odot}$ and the other colours represent intermediate initial masses, increasing from the bottom to the top.}
    \label{fig:massecolorate}
\end{figure}

To further investigate this matter, we chose to mark discs with the same initial mass and follow their evolution through time in order to be able to determine which discs are removed from our sample. This is shown in Fig. \ref{fig:massecolorate}: from $t = 5 \times 10^5$ yr to $5 \times 10^6$ yr the number of discs in the sample decreases, and the discs still present in the sample are the ones with higher initial masses.

This behaviour can be easily understood quantitatively based on the analysis of \citet{ClarkeGendrinSoto2001}, who predict, based on the self-similar evolution, what is the mass $M_{\rm lo}$ that is left-over in the outer disc at the time $t_{\rm w}$ at which the accretion rate through the disc equals the wind outflow rate. Combining Eqs. (3), (13) and (17) in \citet{ClarkeGendrinSoto2001}, one obtains
\begin{equation}
    M_{\rm lo} = 2\dot{M}_{\rm w}t_{\rm w}.
\end{equation}
Now, the knee in the isochrone occurs for those discs for which $t_{\rm w}$ is equal to the instantaneous age of the population $t$ and therefore the disc mass corresponding to the knee, $M_{\rm k}$, is the left-over mass evaluated at $t_{\rm w}=t$:
\begin{equation}
    M_{\rm k} = 2\dot{M}_{\rm w}t,
    \label{eq:kneemass}
\end{equation}
which explains why the knee moves to higher mass for both increasing age and increasing wind mass loss rate. 

\begin{figure}
	\includegraphics[width=\columnwidth]{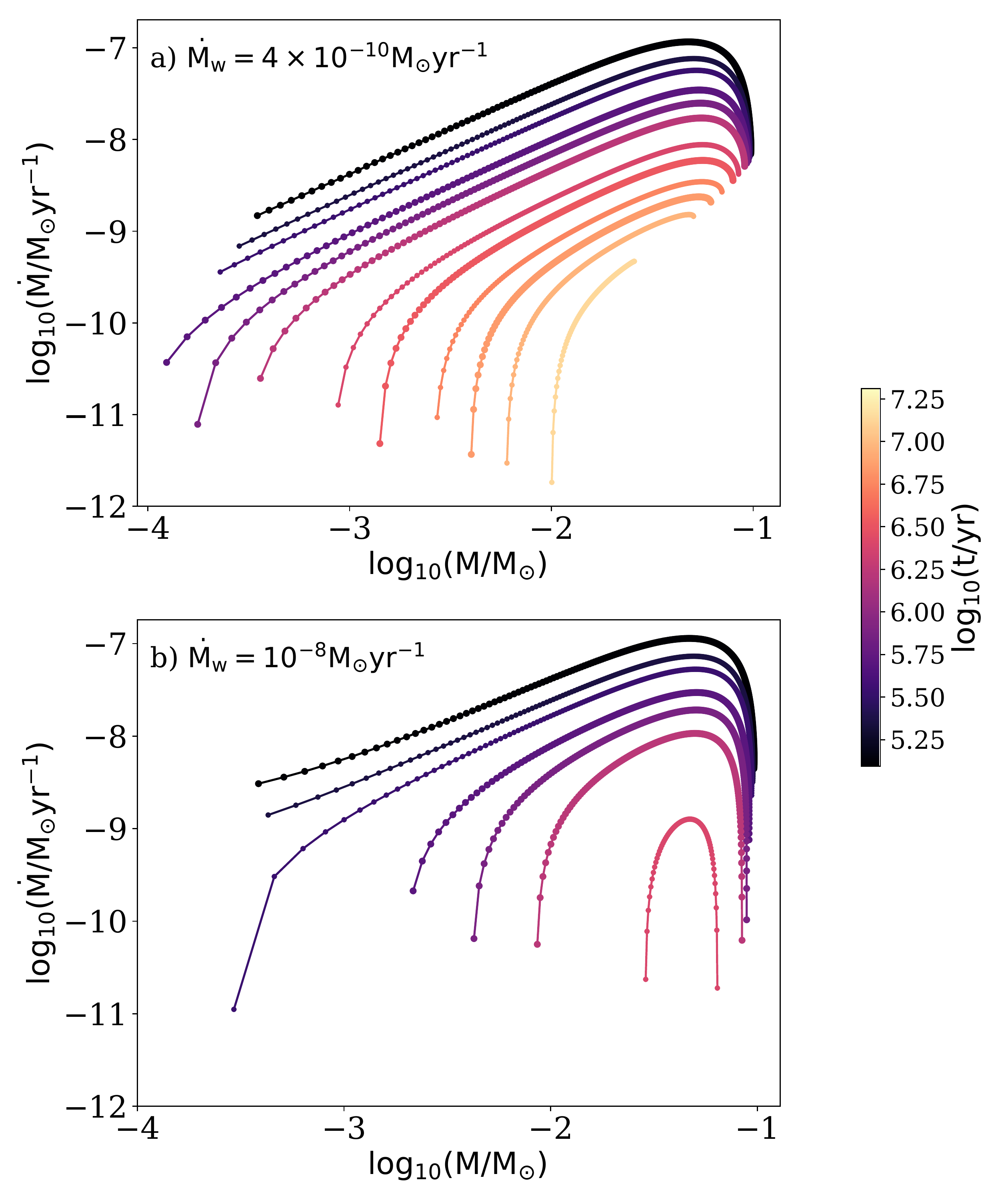}
    \caption{Isochrones for a disc population subject to photoevaporation implemented following \citet{ClarkeGendrinSoto2001} (top) and \citet{Owen2012} (bottom). These plot were obtained setting $M_0 = 10^{-1}$ M$_{\odot}$.}
    \label{fig:isocrone_tanti_tempi}
\end{figure}

To further investigate this point, we restricted the sample fixing also the value of the initial disc mass to $M_0 = 10^{-1} \text{M}_{\odot}$, and let only the values of $R_c$ vary in the ranges described above. Fig. \ref{fig:isocrone_tanti_tempi} shows the results of these simulations at several ages $t$ for both the cases where $\dot{M}_{\rm w}=4 \times 10^{-10} \rm{M}_{\odot} \rm{yr}^{-1}$ (top) and $\dot{M}_{\rm w}=10^{-8} \rm{M}_{\odot} \rm{yr}^{-1}$ (bottom), which clearly illustrates the increasing trend of $M_{\rm k}$ with time.

\begin{figure}
	\includegraphics[width=\columnwidth]{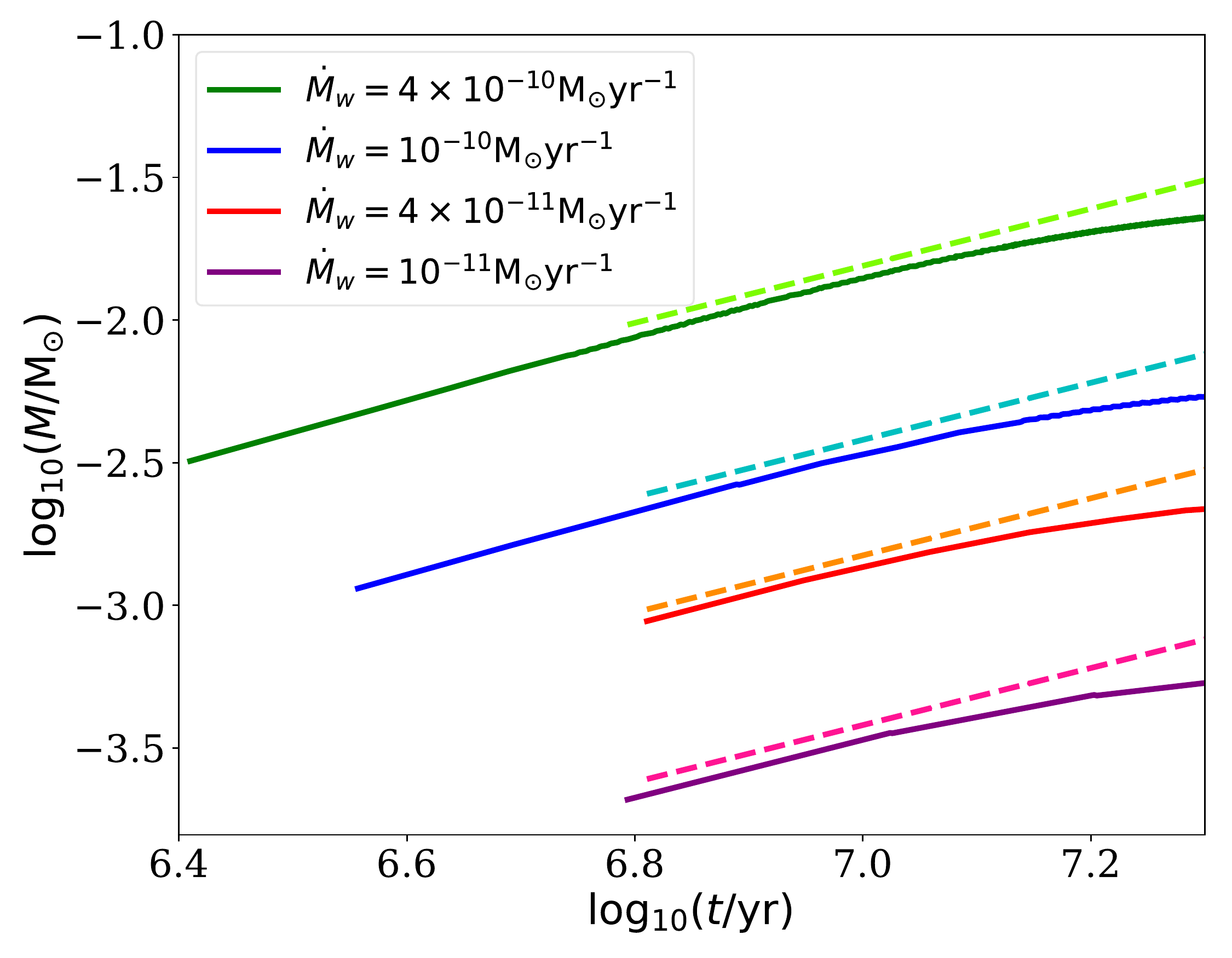}
    \caption{Disc mass for which the accretion rate equals the mass-loss rate as a function of population age. This plot is obtained using four different values of $\dot{\rm{M}}_{\rm{\rm{\rm{\rm{wind}}}}}$, as shown in the legend.}
    \label{fig:asintoti}
\end{figure}

In Fig. \ref{fig:asintoti}, we show the value of
$M_{\rm k}$ (defined here as the mass for which $\dot{M}=\dot{M}_{\rm w}$ at a given age) versus age $t$, for different choices of $\dot{M}_{\rm w}$, along with the theoretical prediction through Eq. \ref{eq:kneemass} (dashed lines). The theoretical prediction is recovered very well, except for very long ages, for which the trend appears to flatten. This deviation is due to the finite extent of our radial grid, so that when the age becomes too long some of the disc mass is lost out of the outer boundary of the grid.

\begin{figure}
	\includegraphics[width=\columnwidth]{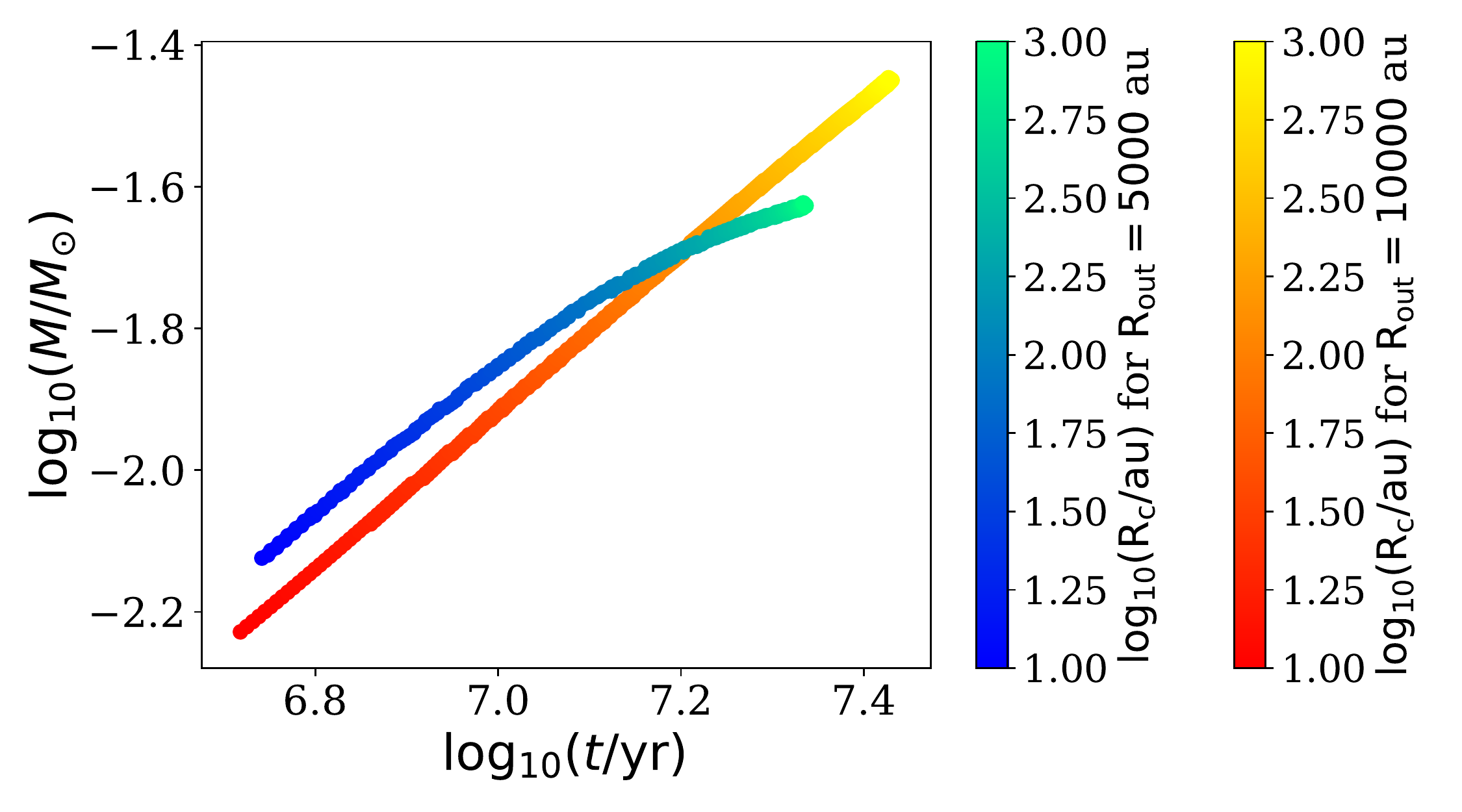}
    \caption{Disc mass for which the accretion rate equals the mass-loss rate as a function of population age. Different colours represent different values of $R_c$, as shown in the colorbar.}
    \label{fig:asintoti1}
\end{figure}

To confirm this interpretation, we tested the effects of the boundary conditions considering the same population evolved on a grid with $R_{\rm{out}}=10000$ au.
The results are shown in Fig. \ref{fig:asintoti1}, where the blue-green points refer to the smaller grid and the red-yellow points refer to the larger grid\footnote{Note, however, that a flattening of the relation is expected to occur when $t\sim t_{\rm w}$ becomes longer than the disc viscous timescale, for which the similarity solutions used by \citet{ClarkeGendrinSoto2001} no longer applies. In this case, the disc mass corresponding to the knee is expected to asymptote at the initial disc mass. We thank Cathie Clarke for pointing this out.}. 

\section{Discussion}\label{sec:discussion}

\begin{figure}
	\includegraphics[width=\columnwidth]{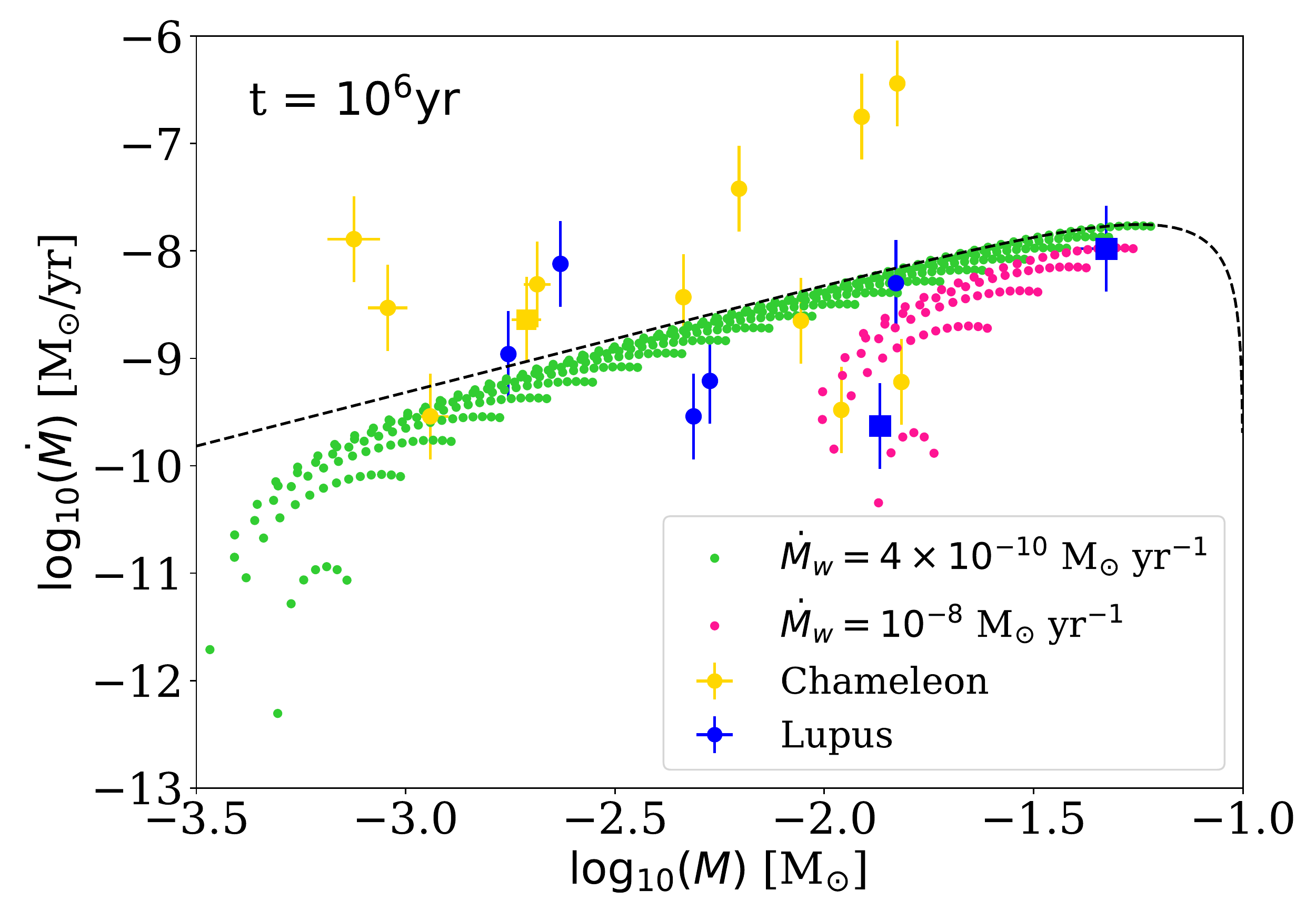}
    \caption{Mass and accretion rates of discs around solar-type stars in the Lupus (blue points) and Chameleon (yellow points) star forming regions; squares indicate objects known as transition discs, with the same colour coding. The green and pink dots show isochrones for a disc population subject to photoevaporation at age $1 \rm{Myr}$ with mass loss rates of $4 \times 10^{-10} \rm{M}_{\odot} \rm{yr}^{-1}$ (green dots) and $10^{-8} \rm{M}_{\odot} \rm{yr}^{-1}$ (pink dots). The dotted black line shows the analytical isochrone at 1 Myr.}
    \label{fig:confronto_dati}
\end{figure}

Recent surveys of large samples of discs with measured mass \citep{Ansdell2016,Ansdell2017,Barenfeld16,Pascucci2016,Cox2017,Cazzoletti2019} and mass and accretion rates \citep{Manara2016a,Mulders2017} are the prime observational test for our models. However, we cannot make a detailed comparison as yet, because here we have considered a limited set of system parameters, and in particular we have only considered a single value for the stellar mass (i.e., 1$ \rm{M}_{\odot}$). In order to properly compare our models to observations we would need to perform an appropriate Montecarlo sampling of the disc initial conditions in terms of disc mass, outer radius and photoevaporation rate. However, we defer this to a future work.

A first, preliminary comparison can be done for the samples of objects collected in the Lupus and Chameleon star forming region, by limiting ourselves to consider only solar-type stars. We have thus selected stars in the mass range $[0.7-1.3] \rm{M}_{\odot}$ within the Lupus and Chameleon samples. We plot the observational data in Fig. \ref{fig:confronto_dati}  (blue points: Lupus; yellow points: Chameleon) along with two sets of models at an age of 1 Myr, assuming a mass loss rate of $\dot{M}_{\rm w}=4~10^{-10} \rm{M}_{\odot}$/yr (green points) and $\dot{M}_{\rm w}=10^{-8} \rm{M}_{\odot}$/yr (pink points) (see section 2.2). The observed data show a much larger scatter than our models. However, note that we have not sampled the parameter space in a statistical sense, and we expect that a broader sampling in terms of initial disc mass and radius can broaden our model distribution (see e.g. \citealt{Lodato2017}). The interesting thing to notice here is that while for $\dot{M}_{\rm w}=4~10^{-10} \rm{M}_{\odot}$/yr our models account for the survival of discs with masses $M\lesssim 10^{-3} \rm{M}_{\odot}$, as observed, these are completely removed if the outflow rate is higher.

For the older (10 Myr) Upper Sco region a similar comparison is more difficult, because the observed stars in this sample are predominantly low-mass stars, with only a few reaching $M_\star\approx 1 \rm{M}_{\odot}$. Note that no accretion rate is available in the region, so that the comparison with our models can only be done on the disc masses. At face value, the average disc mass in Upper Sco is lower than in Lupus and Chameleon \citep{Barenfeld16}, apparently in contraddiction with our predictions. However, this could be simply because of the lower stellar masses of the region, since the photoevaporation rate is a sensitive function of the stellar mass. Additionally, since the mass estimates are based on dust, this apparent decrease in the disc mass could be the coinsequence of a smaller dust-to-gas ratio.

\section{Conclusions}\label{sec:conclusion}

In this paper we studied the isochrones in the plane $M - \dot M$, as introduced by \citet{Lodato2017}, for a population of protoplanetary discs subjected to both viscosity and photoevaporation. \citet{Lodato2017} have already shown that discs are not born on the self-similar branch and need some time to reach that condition: this typically occurs after a viscous time $t_{\nu} \sim 10^7$ Myr, when discs can be considered evolved. If other physical effects apart from viscosity take place, the disc lifetime can be modified: in this paper we focus on the effect of internal photoevaporation.

The main conclusions of our study are the following:
\begin{enumerate}
    \item Photoevaporation induces a `knee' in the isochrone for low disc masses, drastically reducing the accretion rate once it falls below the photoevaporative loss rate, as predicted by \citet{Rosotti2017}.
    \item Such knee implies that, for a photovaporating population, at a given value of the disc mass a large spread in accretion rates is possible, which may explain the large spread in accretion rates at the low end of the disc mass spectrum observed at young ages \citep{Manara2016b,Mulders2017} and in the older Upper Scorpius region (Manara et al. in prep).
    \item The removal of the lower portion of the isochrone produces the counter-intuitive result that more evolved populations (or populations with a higher mass-loss rate) appear more massive. 
    \item The disc mass corresponding to the knee can be easily estimated analytically to be $M_{\rm k}=2\dot{M}_{\rm w}t$, where $\dot{M}_{\rm w}$ is the wind rate and $t$ is the age. Thus, if such a knee is observed in the disc mass vs accretion rate plot in a given survey, one could directly estimate the typical photoevaporation rate of that sample. We note that in the Lupus and Chamaleon surveys (see, for example, \citealt{Manara2016a}) no evidence for such a knee is present, down to disc masses of the order of $10^{-4} \rm{M}_{\odot}$ (if one assumes the disc mass to simply scale with dust mass, with a gas/dust ratio of 100). However, this value is the level at which the sensitivity of the survey drops, and many upper limits on the disc masses are present. Assuming an age of $\sim 1$ Myr, this would imply a photoevaporation rate lower than $\approx 10^{-10} \rm{M}_{\odot} \rm{yr}^{-1}$. Deeper surveys are needed to confirm this result.
\end{enumerate}

We caution that in this paper we have only focused on a single value of the stellar mass, taken to be equal to $1 \rm{M}_{\odot}$. Obviously, the photoevaporation rate is expected to strongly depend on stellar mass. For example, the low disc mass end of the Lupus and Chamaleon sample is dominated by discs around low mass stars, with  $M_\star\approx 0.1-0.2 \rm{M}_{\odot}$. 
For a proper comparison with observations one should therefore simulate a population of discs with varying stellar masses and with an appropriate Montecarlo sampling of the disc initial conditions in terms of disc mass, outer radius and photoevaporation rate. We postpone such an analysis to a later investigation.

\section*{Acknowledgements}

We thank the referee, Richard Parker, for his reading of the manuscript and his review. This work and CT have been supported by the project PRIN INAF 2016 The Cradle of Life - GENESIS-SKA (General Conditions in Early Planetary Systems for the rise of life with SKA).
CT acknowledges financial support provided by the Italian Ministry of Education, Universities and Research through the grant Progetti Premiali 2012 – iALMA (CUP C52I13000140001).
This project and GL have received funding from the European Union’s Horizon 2020 research and innovation programme under the Marie Sklodowska-Curie grant agreement No 823823 (DUSTBUSTERS RISE project).
CFM acknowledges an ESO fellowship and
support by the Deutsche Forschungs-Gemeinschaft (DFG, German Research Foundation) - Ref no. FOR 2634/1 TE 1024/1-1. This work is part of the research programme VENI with project number 016.Veni.192.233, which is (partly) financed by the Dutch Research Council (NWO). This work made use of the Python packages Numpy and matplotlib. 




\bibliographystyle{mnras}
\bibliography{mnras_template} 




\appendix




\bsp	
\label{lastpage}
\end{document}